\documentclass[fleqn,10pt]{wlscirep}
\usepackage[utf8]{inputenc}
\usepackage[T1]{fontenc}
\usepackage{lineno}
\usepackage{multirow}
\usepackage{caption}
\usepackage{afterpage}
\usepackage{placeins}

\title{A multicentric dataset for training and benchmarking breast cancer segmentation in H\&E slides}

\author[1]{Carlijn Lems}
\author[1]{Leslie Tessier}
\author[1]{John-Melle Bokhorst}
\author[1]{Mart van Rijthoven}
\author[1]{Witali Aswolinskiy}
\author[2,3]{Matteo Pozzi}
\author[4,5]{Natalie Klubickova}
\author[6]{Suzanne Dintzis}
\author[7]{Michela Campora}
\author[1,8]{Maschenka Balkenhol}
\author[1]{Peter Bult}
\author[1]{Joey Spronck}
\author[7]{Thomas Detone}
\author[7,9]{Mattia Barbareschi}
\author[10]{Enrico Munari}
\author[11]{Giuseppe Bogina}
\author[12,13,14]{Jelle Wesseling}
\author[12]{Esther H. Lips}
\author[1]{Francesco Ciompi}
\author[1]{Frédérique Meeuwsen}
\author[1,15]{Jeroen van der Laak}
\affil[1]{Department of Pathology, Radboud University Medical Center, Nijmegen, The Netherlands}
\affil[2]{Fondazione Bruno Kessler, Trento, Italy}
\affil[3]{Department of Cellular, Computational and Integrative Biology, University of Trento, Italy}
\affil[4]{Biopticka Laboratory Ltd., Pilsen, Czech Republic}
\affil[5]{Department of Pathology, Faculty of Medicine in Pilsen, Charles University, Pilsen, Czech Republic}
\affil[6]{University of Washington Medical Center, Seattle, Washington, United States}
\affil[7]{Department of Surgical Pathology, Santa Chiara Hospital, APSS, Trento, Italy}
\affil[8]{Canisius Wilhelmina Ziekenhuis, Nijmegen, The Netherlands}
\affil[9]{CISMed - Centre for Medical Sciences, University of Trento, Trento, Italy}
\affil[10]{Pathology Unit, University and Hospital Trust of Verona, Verona, Italy}
\affil[11]{IRCCS Sacro Cuore Don Calabria Hospital, Negrar di Valpolicella, Verona, Italy}
\affil[12]{Division of Molecular Pathology, Netherlands Cancer Institute, Amsterdam, The Netherlands}
\affil[13]{Department of Pathology, Netherlands Cancer Institute - Antoni van Leeuwenhoek Hospital, Amsterdam, The Netherlands}
\affil[14]{Department of Pathology, Leiden University Medical Center, Leiden, The Netherlands}
\affil[15]{Center for Medical Image Science and Visualization, Linköping University, Linköping, Sweden}
\affil[*]{corresponding author: Carlijn Lems (carlijn.lems@radboudumc.nl)}

\begin{abstract}
Automated semantic segmentation of whole-slide images (WSIs) stained with hematoxylin and eosin (H\&E) is essential for large-scale artificial intelligence-based biomarker analysis in breast cancer.
However, existing public datasets for breast cancer segmentation lack the morphological diversity needed to support model generalizability and robust biomarker validation across heterogeneous patient cohorts.
We introduce BrEast cancEr hisTopathoLogy sEgmentation (BEETLE), a dataset for multiclass semantic segmentation of H\&E-stained breast cancer WSIs.
It consists of 587 biopsies and resections from three collaborating clinical centers and two public datasets, digitized using seven scanners, and covers all molecular subtypes and histological grades.
Using diverse annotation strategies, we collected annotations across four classes - invasive epithelium, non-invasive epithelium, necrosis, and other - with particular focus on morphologies underrepresented in existing datasets, such as ductal carcinoma in situ and dispersed lobular tumor cells.
The dataset's diversity and relevance to the rapidly growing field of automated biomarker quantification in breast cancer ensure its high potential for reuse.
Finally, we provide a well-curated, multicentric external evaluation set to enable standardized benchmarking of breast cancer segmentation models.

\end{abstract}
\begin{document}

\flushbottom
\maketitle

\thispagestyle{empty}

\section*{Background \& Summary}

Accurate prognostic and predictive biomarkers are essential for guiding treatment planning for breast cancer patients.
Pathologists primarily assess histological features (e.g., cancer subtype, grade) on hematoxylin and eosin (H\&E)-stained histopathology slides \cite{grading,grading2}, while ongoing research is dedicated to discovering novel biomarkers and characterizing existing ones, such as tumor-infiltrating lymphocytes (TILs) \cite{Loi2019,Denkert2018}.
However, the validation of both established and emerging histopathological biomarkers in large patient cohorts remains limited, largely due to the time-consuming and poorly reproducible nature of biomarker quantification \cite{Rakha2017,Kos2020,VanBockstal2021}.
Advances in artificial intelligence (notably using deep neural networks; so-called deep learning) provide a promising opportunity for automating this process \cite{Mandair2023,McCaffrey2024,Jiang2024,vanRijthoven2024,tiger}.
Since biomarkers are typically assessed in specific tissue regions (e.g., the tumor area), a crucial first step in any automated pipeline is the semantic segmentation of H\&E-stained whole-slide images (WSIs).

The development of robust breast cancer tissue segmentation models relies on access to extensive and diverse annotated training data.
Yet, publicly available datasets lack sufficient representation of various histological and molecular subtypes, resulting in limited variation in histological grade and overall tumor morphology.
For example, the Breast Cancer Semantic Segmentation (BCSS) dataset \cite{tcga-bcss} consists solely of triple-negative breast cancer (TNBC) cases, most of which are high-grade, poorly differentiated tumors of the invasive carcinoma of no special type (NST) subtype.
Although the Tumor InfiltratinG lymphocytes in breast
cancER (TIGER) challenge dataset \cite{tiger-data}, which includes the BCSS dataset, also contains human epidermal
growth factor receptor 2-positive (HER2+) cases, this subtype is also typically high-grade and moderately to poorly differentiated.
Notably, both datasets lack low-grade, well-differentiated tumors of the hormone receptor-positive, HER2-negative (HR+/HER2-) molecular subtype, which accounts for the majority of breast cancer cases.
Moreover, invasive lobular carcinoma (ILC) is underrepresented in both datasets, despite being the second most common invasive histological subtype and exhibiting distinct growth patterns.
Overall, existing datasets for breast cancer segmentation do not sufficiently capture the substantial heterogeneity in breast cancer morphology.

Other public datasets, while sometimes more diverse, are often unsuitable for semantic segmentation, either because the annotations are too coarse \cite{bach,bcnb,bracs} or because they only focus on individual cells or nuclei \cite{post-nat-brca,nucls,janowczyk,kumar,monuseg,peternaylor,Krithiga2020,Ding2023}, thereby omitting larger structures found in breast tissue like ductal carcinoma in situ (DCIS) and reactive features like necrosis.
Segmenting these structures is essential for the automated quantification of many tissue-based biomarkers, including TILs \cite{tils,Amgad2020}.

In this work, we present BrEast cancEr hisTopathoLogy sEgmentation (BEETLE), a multicenter and multiscanner dataset for multiclass semantic segmentation of breast cancer H\&E slides, including both biopsies and surgical resections.
It extends and harmonizes unreleased data from Van Rijthoven et al. \cite{hooknet}, Aswolinskiy et al. \cite{proacting}, and Pozzi et al. \cite{matteo}, along with the TIGER dataset and WSIs from The Cancer Genome Atlas Breast Invasive Carcinoma (TCGA-BRCA) \cite{tcga-brca}.
To our knowledge, our dataset is the first to include the two major histological subtypes of invasive breast cancer, all molecular subtypes, and all histological grades.
The annotations cover four segmentation classes, namely invasive epithelium (i.e., invasive tumor), non-invasive epithelium, necrosis, and other.
We employed a targeted approach to data collection, focusing on breast cancer morphologies that are scarcely represented in existing open datasets, as well as tissue structures that are notoriously difficult to segment, such as DCIS, which is often mistaken for invasive tumor, and individually dispersed tumor cells.
To do so, we used various annotation strategies, including but not limited to manual annotations by pathologists, a custom epithelium segmentation network, and a HoVerNet-based pipeline for annotating isolated tumor cells.

Lastly, we provide a carefully curated external evaluation set for benchmarking breast cancer segmentation models, collected from three clinical centers and digitized using three different scanners.
This set includes 170 densely annotated regions of interest (ROIs) from 54 WSIs that capture much of the breast cancer heterogeneity encountered in clinical diagnostics.
While the images are publicly available on Zenodo, the corresponding annotations are sequestered on the Grand Challenge platform, where submissions are ranked on a public leaderboard.
This setup enables standardized, comparable benchmarking of novel methods, thereby driving advances in breast cancer segmentation.

\section*{Methods}

This section outlines the procedures for case collection, histological preparation, and digitization within our dataset.
We then describe the annotation process, followed by our approach to dataset validation through the training and evaluation of a deep learning model.

\subsection*{Case collection, preparation, and digitization}

We collected H\&E-stained cases from six clinical centers for (1) the development set, (2) the external evaluation set, and (3) the development of an H\&E epithelium segmentation model to facilitate the annotation process.
For the development set, we collected cases from the Radboud University Medical Center (RUMC, Nijmegen, the Netherlands), the Netherlands Cancer Institute (NKI, Amsterdam, the Netherlands), and the Santa Chiara Hospital (SCH, Trento, Italy).
For the evaluation set, we collected cases from the Biopticka Laboratory Ltd. (Biopticka, Pilsen, Czech Republic), the University of Washington Medical Center (UW Medicine, Seattle, WA, USA), and the IRCCS Sacro Cuore Don Calabria Hospital (SCDC, Verona, Italy).
For H\&E epithelium segmentation model development, we collected six H\&E-stained cases from the RUMC, which we digitized, restained with a cytokeratin 8-18 (CK8-18) immunohistochemistry (IHC) marker to highlight epithelium, and rescanned.
CK8-18 is exclusively expressed in epithelial tissue.
Afterwards, each image pair (H\&E/IHC) was co-registered using an existing registration algorithm \cite{Lotz2016}.
The use of cases for this study was approved by the institutional review boards of the centers (RUMC reference 2024-17449; NKI reference IRBdm24-329; SCH reference A1045; Biopticka reference 240611; UW Medicine reference MOD00021161; SCDC reference 25046).
All slides were prepared and digitized in the originating clinical center using different scanners: RUMC (3DHISTECH Pannoramic P250 Flash II and Pannoramic 1000), NKI (Leica Biosystems Aperio AT2), SCH (3DHISTECH Pannoramic 250 Flash III and Leica Biosystems Aperio GT 450 DX), Biopticka (Hamamatsu NanoZoomer S360 and Leica Biosystems Aperio GT 450), UW Medicine (Leica Biosystems Aperio GT 450), and SCDC (Roche Ventana DP 200).
All images were subsequently converted to a standard tagged image file format (TIFF) at 0.5 µm/pixel (roughly corresponding to a 200$\times$ microscope magnification).

For the development set, we also included cases from two public datasets: the TIGER WSIROIS \cite{tiger-data} and TCGA-BRCA \cite{tcga-brca} datasets.
Four slides from the TIGER dataset were excluded following review by a pathology resident (L.T.) due to image quality issues (partial blurriness) or atypical histology (e.g., fibroepithelial/fibrosarcomatous lesions, squamous differentiation/metaplasia).

For the development and evaluation sets, we collected biopsies and resections across all histological grades and molecular subtypes of breast cancer.
The development set consists primarily of NST, ILC, and DCIS cases, along with a small subset of cases from other histological breast cancer types (e.g., cribriform, tubular, mixed).
For evaluation, we collected only NST and ILC cases, ensuring an equal distribution of the TNBC, HER2+, and HR+/HER2- molecular subtypes and histological grades 1-3.

\subsection*{Annotation methods}

We defined four annotation classes: invasive epithelium, non-invasive epithelium, necrosis, and other.
The non-invasive epithelium class primarily comprises healthy glands and DCIS but also includes other non-invasive epithelial morphologies such as lobular carcinoma in situ (LCIS), atypical ductal hyperplasia, and apocrine metaplasia. 
Glandular secretions and lumen were included in the non-invasive epithelium annotations unless the area was sufficiently large that, during model training, a patch could be sampled entirely within the lumen, without including any epithelium.
In our case, this threshold corresponded to an area larger than approximately 256$\times$256 µm, and such regions were annotated as other.
Necrosis and calcifications were consistently annotated as necrosis and other, respectively.
The `other' class includes various breast tissue structures, such as stroma, inflammatory areas, adipose tissue, erythrocytes, skin, and background.

Our dataset was annotated by a team of pathologists, pathology residents, and trained research assistants working under their supervision.
Specifically, five board-certified pathologists (S.D., M.C., M.Bal., P.B., F.M.), two pathology residents (L.T., N.K.), and seven trained research assistants (C.L., R.S., S.v.d.B, F.H., T.S., G.S.d.S, M.v.d.W) contributed to the annotation process.
All manual annotations were made using the open-source software ASAP \cite{asap} or QuPath \cite{qupath}.
We describe the annotation workflow in detail in the following sections.

\subsubsection*{Development set}

The development set includes both dense (i.e., all pixels are labeled) and sparse annotations (i.e., only a subset of pixels are labeled, with unlabeled pixels remaining undefined).
In our dataset, dense annotations are confined to predefined ROIs, whereas sparse annotations consist of scattered labels of individual structures across the WSI without ROI definitions.
The development set initially contained 357 WSIs with sparse manual annotations, which were then combined with the TIGER training set \cite{tiger-data} to form our initial dataset.
The TIGER dataset uses a more granular set of seven class labels, which we remapped to align with our four defined classes by merging in-situ tumor and healthy glands into non-invasive epithelium, and grouping tumor-associated stroma, inflamed stroma, and rest into the `other' class.
TIGER does not separately label necrosis or calcifications within in-situ lesions, instead including them as part of the in-situ tumor class.
To ensure consistency with our class labels, we manually annotated necrosis and calcifications within DCIS lesions in ROIs from the RUMC and the Jules Bordet Institute.
For the TCGA-BRCA subset, we used the existing necrosis annotations within DCIS provided in the original BCSS dataset \cite{tcga-bcss} and checked the ROIs to confirm that no calcifications were present.

\begin{figure}[p]
    \centering
    \includegraphics[width=1.0\linewidth]{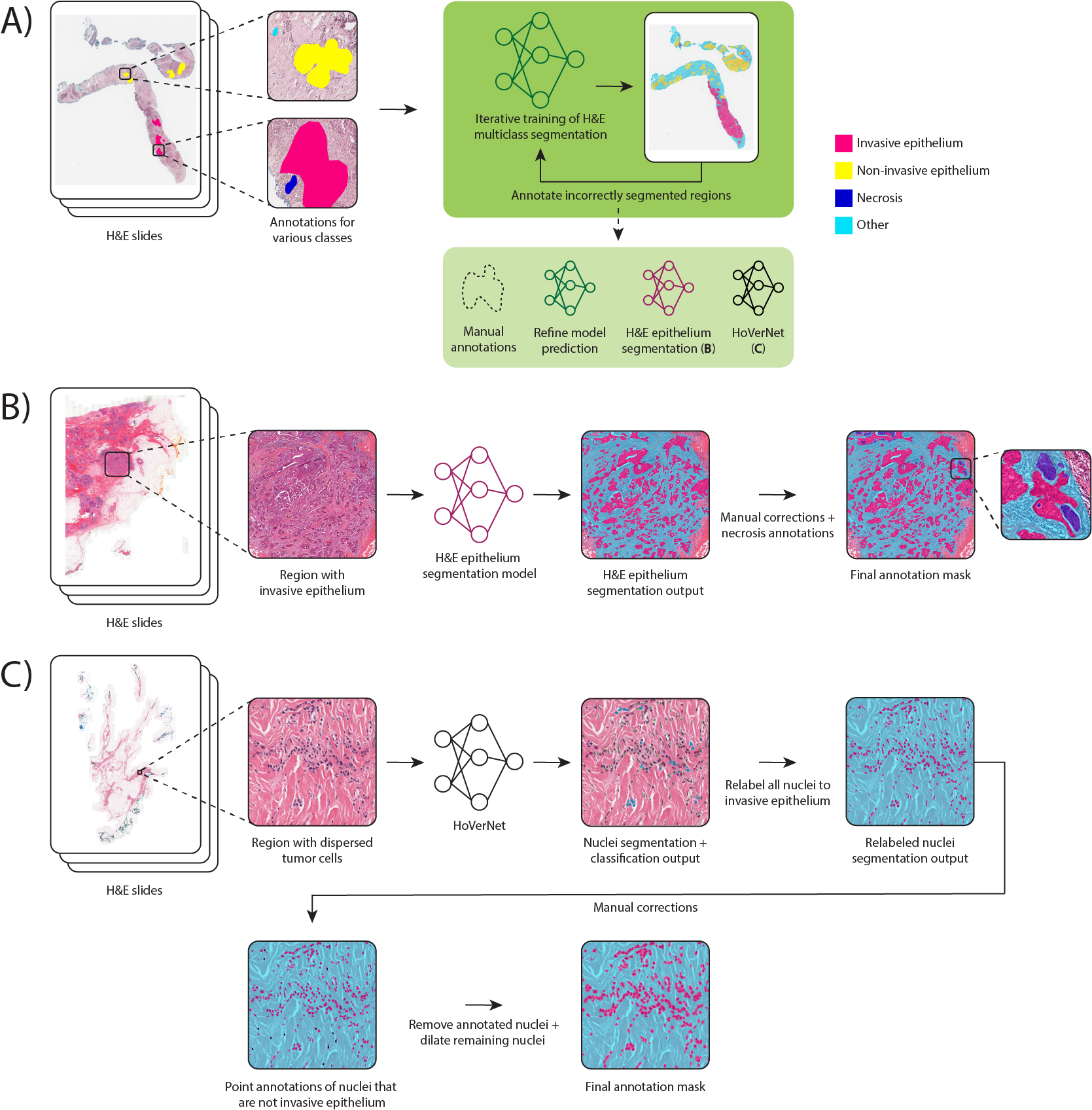}
    \caption{Overview of our annotation methods, which leverage deep learning models for efficient and precise annotations.
    (A) illustrates our iterative workflow for creating new annotations by training an H\&E multiclass segmentation model, applying it to the development set, and refining it through targeted annotation of incorrectly segmented regions.
    (B) and (C) depict pipelines for generating annotations using a custom H\&E epithelium segmentation model and the HoVerNet nuclei segmentation model, respectively.
    }
    \label{fig:anno-methods}
\end{figure}

We subsequently expanded our dataset by training a multiclass segmentation network on our initial dataset, followed by iterative manual hard-negative mining (MHNM) (Fig. \ref{fig:anno-methods}A).
We trained the segmentation model using the \textit{nnU-Net-for-pathology} framework \cite{nnunet,joey}, which trains an ensemble of five U-Net models (see the \textit{Validation methods} section for details).
After each training iteration, we applied the model to the development set at the whole-slide level, selectively annotated incorrectly segmented regions using a combination of dense and sparse annotations, and incorporated them into the development set for the next iteration.
We focused on annotating specific growth patterns and morphologies that were previously underrepresented, such as individual lobular tumor cells, and added a total of 39 new WSIs during this process to enhance dataset diversity.
Furthermore, recognizing that certain classes, such as necrosis and non-invasive epithelium, commonly co-occurred in our initial dataset and could introduce bias, we selectively annotated additional regions to ensure a more balanced and representative development set.
During the MHNM process, we primarily annotated non-TIGER WSIs, annotating new regions in only one slide of the TIGER training set.
We repeated iterations until no further visual improvement was observed.

Incorrectly segmented regions were annotated manually or by merging the model predictions with sparse manual corrections, thereby refining the output.
In some cases, annotators used additional tools to support the annotation process, namely a custom epithelium segmentation network (Fig. \ref{fig:anno-methods}B; see the \textit{Epithelium segmentation network} section for details) or a publicly available nuclei segmentation and classification network \cite{hovernet} (Fig. \ref{fig:anno-methods}C; see the \textit{HoVerNet} section for details).
Depending on the region, annotators selected the most suitable tool to facilitate the annotation process.

\subsubsection*{External evaluation set}

The external evaluation set was annotated either manually or with assistance from the nuclei segmentation network.
ROIs were defined as rectangular areas of approximately 500×500 µm or equivalent, selected to represent the diverse morphologies observed in breast cancer H\&E-stained slides.
For each slide, up to four ROIs were exhaustively annotated, selected to ensure representation of: (1) invasive tumor and tumor-associated stroma; (2) healthy non-tumor tissue, such as stroma, glands, or adipose tissue; (3) potentially challenging regions for segmentation models, including necrosis, inflammation, pigment, or benign lesions (e.g., fibroadenoma, apocrine metaplasia); and (4) DCIS or LCIS.

\subsubsection*{Epithelium segmentation network}

\begin{figure}[hbt!]
    \centering
    \includegraphics[width=1.0\linewidth]{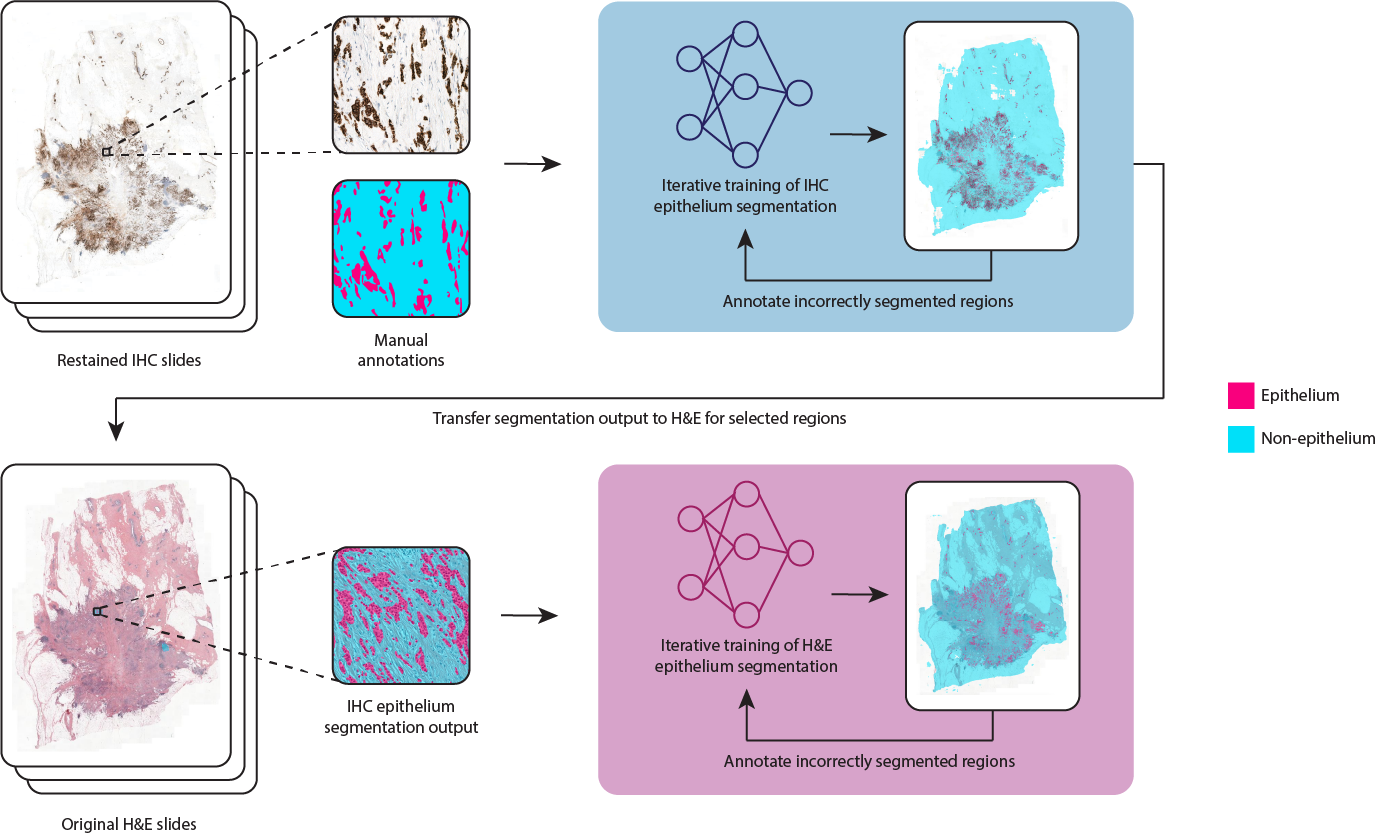}
    \caption{Overview of the two-phase development of a custom H\&E epithelium segmentation network.
    Phase one involved developing a model to segment epithelium in IHC-stained slides.
    Segmentation outputs from selected IHC regions were then transferred to H\&E slides to serve as an initial dataset for developing an H\&E-specific model in phase two.
    Both phases used an iterative cycle of model training, inference, and improvement through annotation of regions with segmentation errors.
    }    
    \label{fig:epi-seg}
\end{figure}

To facilitate efficient and precise annotation of epithelium in invasive tumor regions, we used a two-phase approach to develop an algorithm for segmenting epithelium versus non-epithelial tissue in H\&E-stained breast cancer slides (Fig. \ref{fig:epi-seg}).
In both phases, we trained standard U-Nets \cite{unet,qubvel}, which proved sufficient for the annotation tasks.
We did not use the nnU-Net framework, as it would have introduced unnecessary complexity.
In phase one, we trained a U-Net for IHC epithelium segmentation using six CK8-18-stained slides annotated by a trained researcher (J.M.B.) guided by the IHC signal.
Annotations included dense ROIs containing both epithelium and adjacent non-epithelium, along with sparse annotations of non-epithelial tissue.
We applied MHNM iteratively, applying the model to the WSIs after each training iteration to identify and annotate incorrectly segmented regions as difficult regions for the corresponding class.
New annotations were generated either fully manually or by merging model predictions with sparse corrections in selected regions.
In subsequent iterations, we treated difficult and non-difficult regions of a given class as separate classes with equal sampling probability.
As long as difficult regions were underrepresented, this effectively oversampled them, prompting the model to focus on these challenging areas.
We continued iterative MHNM until no further visual improvement was observed.
In phase two, we applied the final IHC model to all six WSIs, selected several regions, and transferred the network output to co-registered H\&E-stained slides to use as a reference standard for training a second U-Net on H\&E.
We performed MHNM iteratively, without distinguishing difficult from non-difficult regions, until no further visual improvement was observed.
Manual H\&E annotations were performed by a trained researcher (J.M.B.) and a pathology resident (L.T.).
Results from both phases were assessed qualitatively.

The IHC and H\&E networks had a MobileNetV2 \cite{mobilenetv2} and a ResNet50 \cite{resnet50} backbone, respectively, and were initialized with pretrained weights from the ImageNet dataset \cite{imagenet}.
Both models were trained on RGB patches of 512$\times$512 pixels at 1.0 µm/pixel, which were randomly sampled from the annotated regions with an equal probability for each class.
The applied data augmentation techniques included random flipping, blurring, compression, elastic transformations, and HED augmentation.
In both phases, model performance on the validation set was monitored during training using the Dice coefficient.
The networks were optimized using the Lovasz loss \cite{lovasz}, a batch size of 25, and an initial learning rate of 1e-4, which was halved after every 25 epochs if there was no decrease in the validation loss.
The IHC and H\&E networks were trained for a maximum of 50 and 100 epochs, respectively, with 1250 iterations per epoch, and training was stopped if validation performance did not increase for 10 epochs.

For our development set, we generated annotations by applying the final H\&E epithelium segmentation network to breast cancer slides (Fig. \ref{fig:anno-methods}B).
Subsequently, a (resident) pathologist or trained research assistant selected regions with invasive epithelium and tumor-associated stroma, corrected the segmentation output if needed, and manually annotated necrosis when present. 

\subsubsection*{HoVerNet}

A characteristic growth pattern of ILC involves individual, dispersed tumor cells, which are time-consuming to annotate manually with high precision.
To address this, we developed an annotation pipeline that relies on the nuclei segmentation and classification network HoVerNet \cite{hovernet} (Fig. \ref{fig:anno-methods}C).
The pipeline consisted of the following steps:
(1) ROI selection: A pathologist manually selected rectangular ROIs of isolated tumor cells within stroma.
For the development set, regions were selected where preliminary versions of the multiclass segmentation model missed the majority of tumor cells by segmenting them as other.
(2) HoVerNet inference: We applied HoVerNet to the ROIs at 0.25 µm/pixel and labeled all segmented nuclei as invasive epithelium, irrespective of HoVerNet's own classification output.
(3) Manual corrections:
A pathologist reviewed the segmentation output and point-annotated nuclei that were not invasive epithelium, specifying their correct class.
Furthermore, the pathologist annotated necrosis and non-invasive epithelium with polygons.
We then combined these annotations with the HoVerNet segmentation output to obtain refined annotation masks.
(4) Nuclei dilation: Since HoVerNet segments nuclei rather than entire cells, we slightly dilated the nuclei boundaries to approximate whole-cell coverage.
We selected the dilation factor based on visual assessment and applied it uniformly across all slides.

\subsection*{Validation methods}

For technical validation of the BEETLE dataset, we trained a deep learning model for multiclass tissue segmentation on the full development set.
For model training, we used the nnU-Net framework \cite{nnunet}, a self-configuring U-Net-based approach that automatically selects optimal hyperparameters for a given development set.
Specifically, we employed the pathology-specific adaptation \textit{nnU-Net-for-Pathology} developed by Spronck et al. \cite{joey}, which implements a 5-fold cross-validation approach.
We performed a stratified data split at the patient level based on the presence of annotated invasive breast cancer, and, when applicable, histological breast cancer type (NST, ILC, other, or unknown).
Moreover, we ensured comparable annotation class distributions across all folds.

All networks were trained on RGB patches of 512$\times$512 pixels at 0.5 µm/pixel sampled using a balanced sampling strategy to ensure equal sampling of each annotation class, thereby mitigating class imbalance.
Model performance on the validation folds was monitored during training using the Dice coefficient, and for each fold, the checkpoint with the highest validation Dice coefficient was selected for inference and evaluation.
See the \textit{Code Availability} section for details regarding the publicly available model inference script and model weights.

We evaluated segmentation performance on the development set using 5-fold cross-validation.
For each fold, we applied the model to its corresponding validation set, aggregated the confusion matrices across all annotated regions, and computed the Dice coefficient for each class.
We report the mean and standard deviation of the Dice scores across the five folds.
For evaluating our model's performance on unseen data, we applied the ensemble of five trained models to the experimental and final test sets of the TIGER challenge \cite{tiger} and to our external evaluation set.
For TIGER, we mapped all predictions other than invasive epithelium to the rest class.
For the external evaluation set, we expanded the spatial context of each ROI, allowing predictions to incorporate the neighboring tissue architecture as they would during whole-slide inference using a sliding-window approach.
We then aggregated the confusion matrices across all annotated regions and computed the Dice coefficient for each class, reporting results by center, by histological subtype (for invasive epithelium only), and across the full dataset.

\section*{Data Overview}

\begin{figure}
    \includegraphics[width=1.0\linewidth]{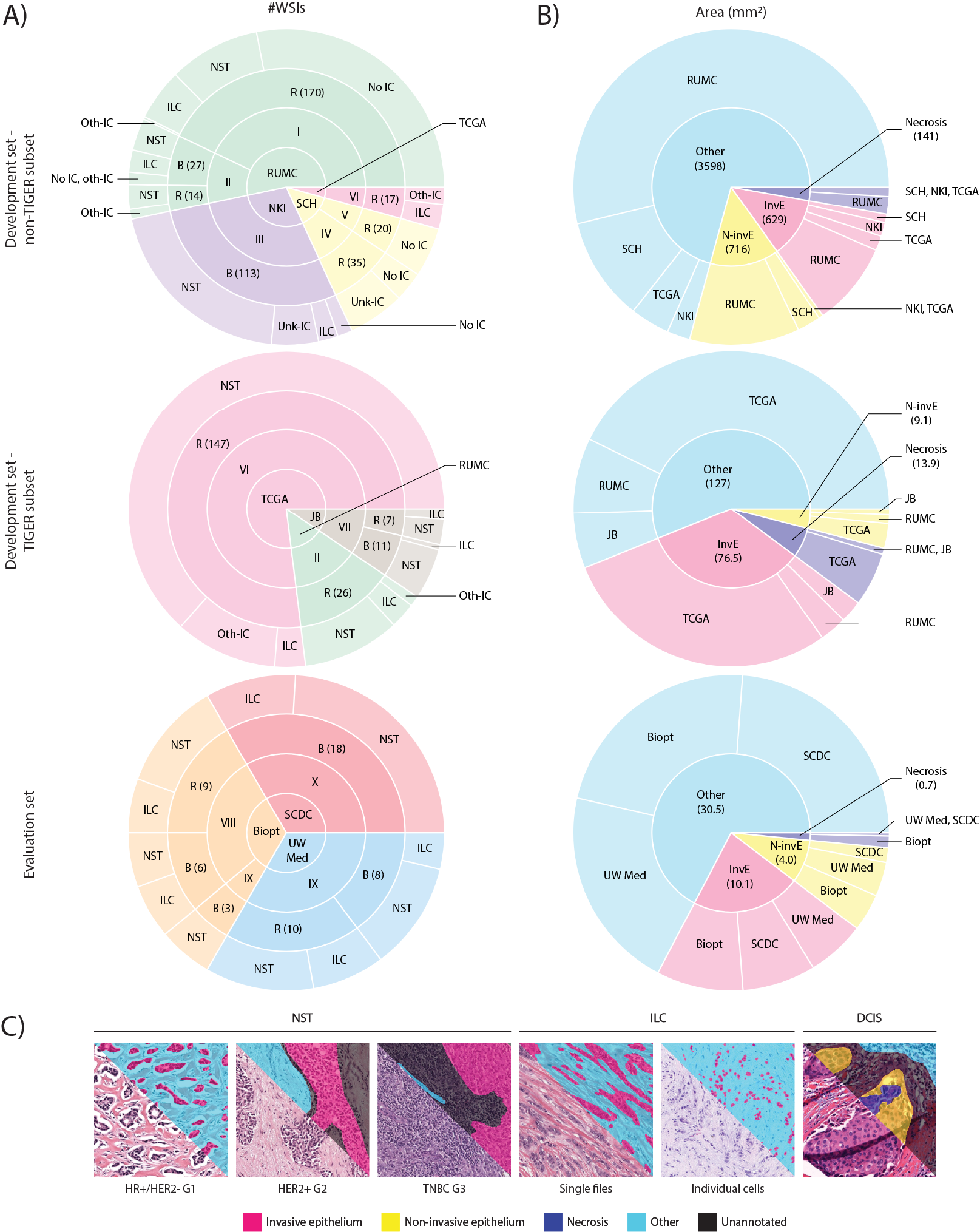}
\end{figure}

\afterpage{%
\begin{center}
\captionof{figure}{
    Overview of the BEETLE dataset.
    (A) displays the distribution of WSIs across clinical centers, scanners, specimen types, and annotated histological breast cancer subtypes for the non-TIGER subset of the development set (top), the modified TIGER training set (middle), and the external evaluation set (bottom).
    `No IC' indicates slides where no invasive carcinoma was annotated, although invasive carcinoma may still be present.
    (B) shows the distribution of annotated area across annotation classes and clinical centers for the same three data subsets.
    (C) provides examples of annotated regions in the development set, highlighting the morphological diversity and staining variation present in the dataset.
    Scanner types: 
    I: 3DHISTECH Pannoramic P250 Flash II; 
    II: 3DHISTECH Pannoramic 1000; 
    III: Leica Biosystems Aperio AT2; 
    IV: 3DHISTECH Pannoramic 250 Flash III; 
    V: Leica Biosystems Aperio GT 450 DX; 
    VI: Leica Biosystems Aperio Scanscope XT; 
    VII: Hamamatsu NanoZoomer 2.0-RS (C10730 series);
    VIII: Hamamatsu NanoZoomer S360; 
    IX: Leica Biosystems Aperio GT 450; 
    X: Roche Ventana DP 200.
    Abbreviations: 
    B: biopsy;
    DCIS: ductal carcinoma in situ;
    G: grade;
    HER2+: human epidermal growth factor receptor 2-positive;
    HR+/HER2-: hormone receptor-positive, HER2-negative;
    IC: invasive carcinoma;
    ILC: invasive lobular carcinoma;
    InvE: invasive epithelium;
    N-invE: non-invasive epithelium;
    NST: no special type;
    Oth-IC: other IC;
    R: resection;
    TNBC: triple-negative breast cancer;
    Unk-IC: unknown IC.}
\label{fig:data}
\end{center}
}

Figure \ref{fig:data} provides an overview of the BEETLE dataset.
Panel A shows the distribution of slides across clinical and technical variables, such as histological subtype and scanner, while Panel B displays the total annotated area, stratified by annotation class and clinical center.
Histological subtypes for all TCGA-BRCA slides were obtained from the 2016 TCGA-BRCA histologic type annotations provided by Thennavan et al.\cite{Thennavan2021}.
The development set comprises 5567 mm$^2$ of annotated tissue across 587 slides from 527 patients, while the external evaluation set includes a total annotated area of 45 mm$^2$ of tissue across 54 slides from 54 patients.
Panel C presents representative examples illustrating the diversity of annotated cases, which cover all molecular subtypes, histological grades, and various growth patterns.

\section*{Technical Validation}

\begin{figure}
    \centering
    \includegraphics[width=1.0\linewidth]{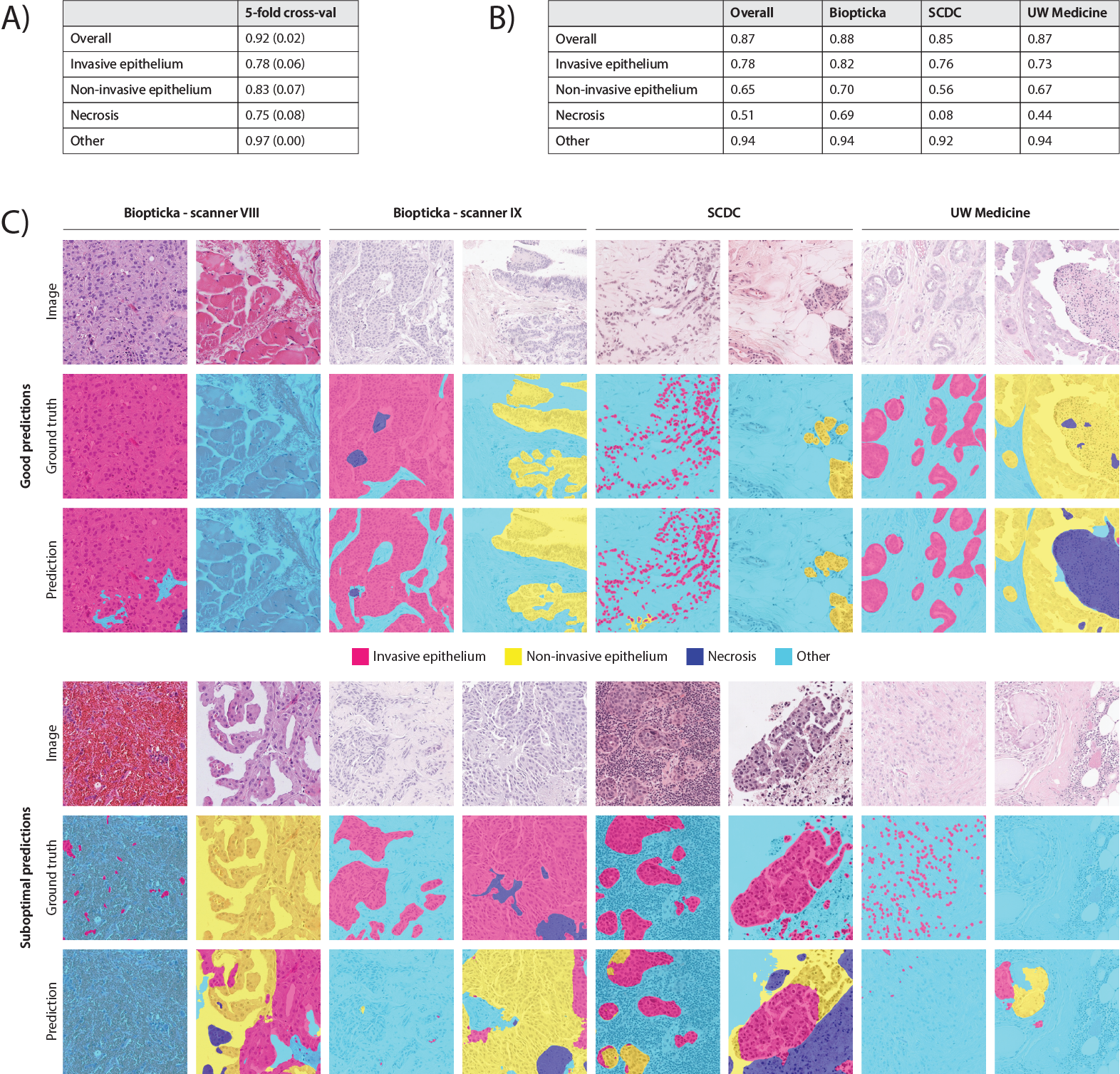}
    \caption{Summary of the technical validation of the BEETLE dataset, demonstrated by the segmentation performance of a deep learning model trained on our development set.
    (A) reports 5-fold cross-validation Dice coefficients (mean $\pm$ standard deviation) for each annotation class.
    (B) presents Dice coefficients for the multicentric external evaluation set for each annotation class and each center, and aggregated across the full dataset.
    (C) provides visual examples of model performance on the external evaluation set.
    Scanner types:
    VIII: Hamamatsu NanoZoomer S360; 
    IX: Leica Biosystems Aperio GT 450.
    }
    \label{fig:tech-val}
\end{figure}

We evaluated the performance of our segmentation model to demonstrate the technical validity of the BEETLE dataset.
The overall model performance is summarized in Fig. \ref{fig:tech-val}, reporting results from 5-fold cross-validation and evaluation on our external multicentric test set.
5-fold cross-validation yielded an overall Dice coefficient of 0.92 (Fig. \ref{fig:tech-val}A).
Class-wise performance was highest for the `other' tissue class (0.97), followed by non-invasive epithelium (0.83), invasive epithelium (0.78), and necrosis (0.75).

The model was also evaluated on the public TIGER challenge experimental and final segmentation test sets \cite{tiger}, achieving invasive tumor Dice coefficients of 0.79 and 0.77, respectively. 
These results are comparable to the highest-ranked submissions on the respective leaderboards, which reported Dice coefficients of 0.79 (experimental) and 0.81 (final) for invasive tumor.
We could only assess invasive tumor performance, as the TIGER test sets can only be evaluated for tumor and stroma, and stroma was not among our segmentation classes.

Evaluation on the multicentric external test set showed a slight decline in performance, with an overall Dice coefficient of 0.87 across all centers (Fig. \ref{fig:tech-val}B).
The invasive epithelium score remained stable at 0.78 and was consistent across histological subtypes (0.77 for NST, 0.78 for ILC), while other classes showed decreased performance: other (0.94), non-invasive epithelium (0.65), and necrosis (0.51).
The lower necrosis score can be partly attributed to class imbalance in the test set, with necrosis comprising less than 2\% of annotated area overall (Fig. \ref{fig:data}B, bottom).
In such cases, even a small number of false positives can considerably reduce the Dice coefficient. 
This effect is further illustrated by center-wise results: the model achieved its highest necrosis Dice score for Biopticka, the center with the largest necrosis proportion (0.69 Dice; 0.03\% necrosis), and the lowest score for SCDC, which had the smallest necrosis proportion (0.08 Dice; <0.01\% necrosis).

Fig. \ref{fig:tech-val}C presents representative examples of segmentation performance.
The model effectively segments invasive epithelium across a variety of morphologies, including high-grade solid tumors, well-differentiated low-grade tumors, and discohesive and single-file patterns of lobular tumor cells.
A frequent source of error is the misclassification of invasive epithelium as non-invasive, particularly in regions lacking sufficient tissue context, such as isolated fragments at biopsy borders or dense tumor regions resembling the core of a DCIS lesion.
However, this misclassification typically affects only a subset of the tumor that is representative of the overall tumor tissue, without clear bias toward factors like tumor grade, and is therefore unlikely to substantially impact the outcome of downstream tasks such as biomarker quantification.
Furthermore, despite diverse training data, the model occasionally performs less on epithelium segmentation in slides with markedly low tissue contrast.

Beyond invasive epithelium, the model typically performs well on common breast tissue structures, including healthy epithelium (lobules and ducts), hyperplasia, adenosis, and DCIS (non-invasive epithelium), as well as stromal components such as blood vessels, muscle, nerves, sclerosis, and inflammation (other) (Fig. \ref{fig:tech-val}C).
Performance decreased on features underrepresented in the training data, such as apocrine metaplasia and granulomatous inflammation, which are not uncommon in breast pathology and thus included in the benchmark to capture tissue variability.
Finally, necrosis segmentation was affected by the occasional misclassification of debris and blood, as well as a tendency to segment not only fully necrotic cells but also early necrotic or injured cells with partially preserved nuclei.

\section*{Data Availability}

The BEETLE dataset is available on Zenodo \cite{beetle}.
The development set includes public WSIs from the TIGER training set, which are available separately via the AWS Open Data Registry \cite{tiger-data}.
Annotations for the external test set are sequestered on the Grand Challenge platform (\url{https://beetle.grand-challenge.org/}) and are not publicly released.

\section*{Code Availability}

The code is available in our GitHub repository (\url{https://github.com/DIAGNijmegen/beetle}).
It includes scripts to download the dataset from Zenodo and to run model inference.

\bibliography{main}

\section*{Acknowledgements}

The authors thank Roan Sherif, Sophie van den Broek, Frederike Haverkamp, Tinka Santing, Gabriel Silva de Souza, and Milly van de Warenburg for their support in annotating the cases.
The authors also thank Alon Vigdorovits for reviewing the qualitative performance of an H\&E multiclass tissue segmentation model trained on an intermediate version of the dataset presented in this work.
We would like to acknowledge the Core Facility Molecular Pathology and Biobanking at the NKI for technical support and assistance in scanning the H\&E slides.
This work was supported by a research grant from the Dutch Cancer Society (KWF, COMMITMENT project number 15386).
Research at the Netherlands Cancer Institute is supported by institutional grants of the Dutch Cancer Society and of the Dutch Ministry of Health, Welfare and Sport.

\section*{Author Contributions}

C.L. wrote the manuscript, designed the experiments, and set up the benchmark on the Grand Challenge platform.
C.L., L.T., M.v.R., W.A., and M.P. oversaw the collection and curation of different subsets of the H\&E multiclass tissue segmentation dataset.
C.L., L.T., N.K., S.D., M.C., M.Bal., P.B., and F.M. annotated or supervised annotations for this dataset.
J.M.B. annotated cases for IHC epithelium segmentation, and L.T. and J.M.B. jointly annotated cases for H\&E epithelium segmentation; they co-developed the corresponding algorithms.
F.M. annotated the external evaluation set.
C.L. and J.M.B. set up the HoVerNet-based segmentation pipeline.
L.T., M.Bal., and F.M. provided histopathological input.
J.S. provided technical advice on training the H\&E multiclass tissue segmentation model.
M.C., T.D., and M.Bar. collected cases from SCH; M.Bal. and P.B. from RUMC; J.W. and E.H.L. from NKI; N.K. from Biopticka; S.D. from UW Medicine; and E.M. and G.B. from SCDC.
F.C. and J.v.d.L. conceived the study, co-designed experiments, and supervised the work.
All authors reviewed the manuscript and approved its contents.

\section*{Competing Interests}

W.A. is currently employed at PAICON GmbH, Germany; the work was conducted prior to this employment.
M.Bal. is medical advisor at Aiosyn BV, the Netherlands.
F.C. is shareholder of Aiosyn BV, the Netherlands.
J.v.d.L. was a member of the advisory boards of Philips, the Netherlands, and ContextVision, Sweden, and received research funding from Philips, the Netherlands, ContextVision, Sweden, and Sectra, Sweden, in the last five years. He is Chief Scientific Officer (CSO) and shareholder of Aiosyn BV, the Netherlands.
All other authors declare no conflict of interest.

\end{document}